\documentclass[twocolumn,preprintnumbers,aps,prl,amsmath,amssymb,superscriptaddress,showpacs]{revtex4}

\usepackage{graphicx}
\usepackage{dcolumn}
\usepackage{bm}
\usepackage{amssymb}
\usepackage[german, english]{babel}
\usepackage{color}
\begin{document}

\title{Magnetically Controlled Exchange Process in an Ultracold Atom-Dimer Mixture}

\author{S. Knoop}\email[Present address: Kirchhoff-Institut f\"ur Physik, Universit\"at Heidelberg, Germany]{}
\author{F. Ferlaino}
\author{M. Berninger}
\author{M. Mark}\email[Present address: Swinburne University of Technology, Melbourne, Australia]{}
\author{H.-C. N\"{a}gerl}\affiliation{Institut f\"ur Experimentalphysik and Zentrum f\"ur Quantenphysik, Universit\"at
 Innsbruck, 6020 Innsbruck, Austria}
\author{R. Grimm}\affiliation{Institut f\"ur Experimentalphysik and Zentrum f\"ur Quantenphysik, Universit\"at Innsbruck, 6020 Innsbruck, Austria}
\affiliation{Institut f\"ur Quantenoptik und Quanteninformation, \"Osterreichische Akademie der Wissenschaften, 6020 Innsbruck, Austria}
\author{J. P. D'Incao}\affiliation{JILA, University of Colorado and NIST, Boulder, Colorado 80309-0440, USA}
\author{B. D. Esry}\affiliation{Department of Physics, Kansas State University, Manhattan, Kansas 66506, USA}

\date{\today}

\begin{abstract}
We report on the observation of an elementary exchange process in an optically trapped ultracold sample of atoms and Feshbach molecules. We can magnetically control the energetic nature of the process and tune it from endoergic to exoergic, enabling the observation of a pronounced threshold behavior. In contrast to relaxation to more deeply bound molecular states, the exchange process does not lead to trap loss. We find excellent agreement between our experimental observations and calculations based on the solutions of three-body Schr\"{o}dinger equation in the adiabatic hyperspherical representation. The high efficiency of the exchange process is explained by the halo character of both the initial and final molecular states.
\end{abstract}


% 31.15.xj, 34.50.-s,

\maketitle

The full control of reactive processes on the quantum level is a major prospect of ultracold chemistry. The strong current efforts to produce samples of trapped ultracold molecules \cite{Krems2009cmt,Danzl2008qgo} are providing experimentalists with new systems, where fundamental questions in this emerging field can be addressed \cite{Carr2009cau}. Molecules can be prepared at ultralow temperatures in a single internal state, i.e.\ within a specific quantum state of its vibrational, rotational or spin quantum numbers. This allows quantum-state selective studies of elastic, inelastic, and even chemically reactive collisions. The application of external electromagnetic fields opens up unique possibilities for controlling the interacting processes, thus leading to a controlled ultracold chemistry \cite{Krems2008ccc}.

Elementary interaction processes in ultracold molecular gases have been studied experimentally, with the observation of intriguing phenomena such as the stability of Feshbach molecules created from fermionic atoms \cite{Strecker2003coa} and the observation of collisional resonances caused by the presence of trimer \cite{Knoop2009ooa} and tetramer states \cite{Chin2005oof}. Inelastic collisions are commonly probed in a trap-loss regime where the products leave the trap following a large release of internal energy; see e.g.\ \cite{Mukaiyama2004dad}. Such trap loss represents a readily detectable signature of the interactions, but it does not provide information on the reaction products. The situation changes, however, when the energies of the initial and final state are nearly degenerate, and the small amount of energy released allow the reaction products to remain in the trap. Such a scenario provides unique access to study the dynamics of the reaction but has, to our knowledge, not been observed yet in ultracold gases.

In this Letter, we report on the observation of a controllable exchange process ${\rm A}_2+{\rm B} \rightarrow {\rm A}+{\rm AB}$ in an ultracold mixture of weakly bound dimers (${\rm A}_2$) and free atoms (B). Our experiments provide a proof-of-principle demonstration of an elementary exchange reaction in the ultracold regime, including the direct observation of A atoms as a reaction product. Magnetic tuning allows to vary the energetic nature from endoergic over resonant to exoergic. We theoretically study the process by numerically solving the three-body Schr\"{o}dinger equation in the adiabatic hyperspherical representation \cite{Suno2002tbr} using a model potential. The calculations highlight the important role of large scattering lengths in the universal halo regime where both the ${\rm A}_2$ and the AB dimers are very weakly bound \cite{Kohler2006poc,Ferlaino2009ufm}. Our results can be interpreted as the first observation of a reaction in the regime of universal interactions \cite{Braaten2006uif}.

The two atomic constituents A and B are represented by Cs atoms in different hyperfine states. More specifically, A represents the lowest hyperfine sublevel $|F=3,m_F=3\rangle$, and B one of the upper $|F=4,m_F\rangle$ hyperfine sublevels, with $m_F= 2$, 3 or 4; the quantum number $F$ denotes the total spin and $m_F$ its projection. The characterization of the weakly bound dimer states according to the atomic hyperfine spin states is possible as the molecular binding energies are much smaller than the atomic hyperfine and Zeeman splitting. Since A and B only differ by their spin state, atom exchange is indistinguishable from spin exchange and the exchange process generally represents a coherent sum of both.

\begin{figure}
\includegraphics[width=8.5cm]{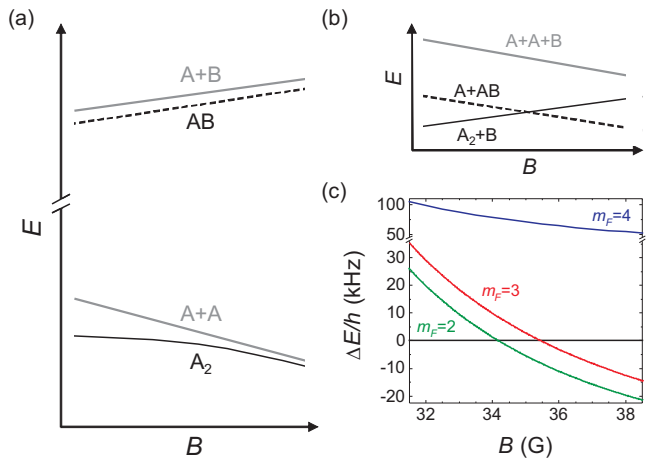}
\caption{(color online) (a) Zeeman diagram of the most weakly bound dimer states A$_2$ and AB below the A+A and A+B dissociation thresholds, respectively; here A and B are two hyperfine sublevels of Cs. (b) Schematic representation of a crossing between the A$_2$+B and A+AB channels. (c) The energy difference $\Delta E$ between the A$_2$+B($m_F$) and A+AB($m_F$) channels, with $m_F=2$, 3 or 4, showing channel crossings around 35\,G for $m_F=2$ and 3.}\label{energydiagram}
\end{figure}

The magnetic field dependencies of the relevant molecular states play a crucial role for the exchange process. In Fig.~\ref{energydiagram}(a) we schematically show the Zeeman diagram of the most weakly bound states A$_2$ and AB below the A+A and A+B dissociation thresholds, respectively. The magnetic field dependence of the A$_2$ binding energy has been studied both experimentally and theoretically \cite{Chin2004pfs,Mark2007sou}. Towards lower magnetic fields the A$_2$ state bends away from the A+A threshold, and its binding energy increases from $h\times30$\,kHz at 45\,G to $h\times150$\,kHz at 30\,G. The AB binding energies are essentially independent of the magnetic field and amount to about $h\times5$\,kHz for $m_F=4$, and $h\times80$\,kHz for $m_F=2$ and $m_F=3$. We obtain these values from scattering length calculations \cite{Julienne2008privat}, using a generalized relation between the scattering length and the binding energy derived from quantum defect theory \cite{Gao2004bea}.

As a result of the different magnetic field dependencies of the A$_2$ and AB binding energies, the A$_2$+B and A+AB channels can cross each other, which is depicted in Fig.~\ref{energydiagram}(b). Such a scenario provides the opportunity to magnetically tune the exchange process into resonance. A quantitative picture is shown in Fig.~\ref{energydiagram}(c), showing the energy difference $\Delta E$ between the A$_2$+B and A+AB channels. The channels cross for $m_F=2$ and 3 around 35\,G, and for higher magnetic fields the exchange process ${\rm A}_2+{\rm B} \rightarrow {\rm A}+{\rm AB}$ is exoergic and thus energetically allowed. For $m_F=4$ the exchange process remains endoergic in the investigated magnetic field range.

The experimental setup has been described earlier \cite{Ferlaino2008cbt,Knoop2009ooa}. In brief, we prepare an ultracold thermal sample of Cs atoms at temperatures of 50-100 nK in state A in a crossed-beam optical dipole trap, with a mean trap frequency of typically 30\,Hz. Feshbach association at a narrow Feshbach resonance at 48\,G results in an A+A$_2$ atom-molecule mixture consisting of $4000$ molecules and $30,000$ atoms. By applying a 3-ms microwave (MW) pulse we transfer the atoms from state A to state B with an efficiency of better than 95$\%$ and without any observable effect on the A$_2$ molecules. After preparation of the B+A$_2$ mixture we ramp to a certain magnetic field and wait for a variable storage time. Then we switch off the trap and let the sample expand before ramping back over the 48-G resonance to dissociate the remaining A$_2$ molecules, after which standard absorption imaging is performed. During expansion we apply the Stern-Gerlach technique to spatially separate the different atomic and molecular states.

\begin{figure}
\includegraphics[width=8.5cm]{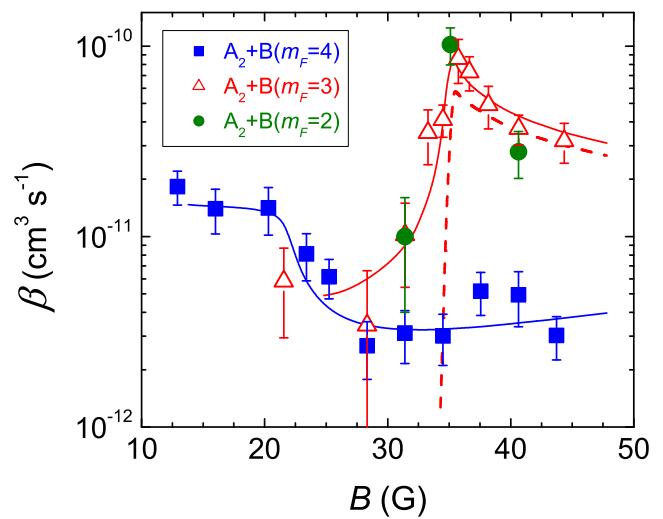}
\caption{(color online) Rate coefficient $\beta$ for inelastic atom-molecule collisions
for the B+A$_2$ mixtures at a temperature of 50(10)~nK as function of the magnetic field $B$, comparing the experimental results (symbols) with the model calculations (lines). The solid curves represent the total A$_2$ loss rate including relaxation to more deeply bound states, whereas the dashed curve shows the contribution of the exchange process. The error bars contain the uncertainties of the trap frequencies and temperature measurements, required to convert the measured particle numbers into densities.}\label{lossrates}
\end{figure}

In a first set of experiments, we measure the atom-molecule inelastic collision rates by recording the time evolution of the number of A$_2$ molecules and B atoms. In order to extract a rate coefficient $\beta$ from the data we set up a loss model, similar to the one described in Ref.~\cite{Knoop2009ooa}. The model includes loss by molecule-molecule collisions \cite{Ferlaino2008cbt} and atomic loss due to two-body hyperfine-changing collisions. We derive an analytic expression, which we fit to the data. We show the results in Fig.~\ref{lossrates}, showing $\beta$ as function of the magnetic field. The measured loss rate coefficient includes all atom-molecule collisions that lead to the loss of A$_2$ molecule from the sample, i.e.\ both the exchange process as well as relaxation to more deeply bound vibrational states.

In parallel, we theoretically determine the loss rates for A$_2$+B collisions. We model the two-body interactions with short-range potentials that have only a few bound states. Their strength is chosen in such a way to reproduce the energy of the weakly bound molecular A$_2$ and AB states at each magnetic field. The results of our calculations are shown in Fig.~\ref{lossrates} for $m_F=3$ and 4. The solid curves display the total loss rate of A$_2$ molecules, including both the exchange process and relaxation to more deeply bound states. The dashed curve shows the contribution that results from the exchange process. Because of the simplicity of the model potential the calculations are not expected to have accurate predictive power regarding the absolute total loss rate. Therefore the calculations are normalized to the experimental data to facilitate the comparison. The comparison between theory and experiment shows excellent agreement in the dependencies of the total loss rates on the magnetic field.

The most striking observation is the resonant enhancement that the rate coefficient $\beta$ shows for both $m_F=2$ and 3 around 35\,G. The calculations show that the resonance is caused by opening up the possibility for the exchange process. Once energetically allowed, it contributes to 80\% of all the inelastic atom-molecule collisions. Within the theoretical model the results for $m_F=2$ (not shown in Fig.~\ref{lossrates}) are exactly the same as for $m_F=3$, but simply shifted in magnetic field by 1.3~G because of the slight difference in AB binding energy.

\begin{figure}
\includegraphics[width=8cm]{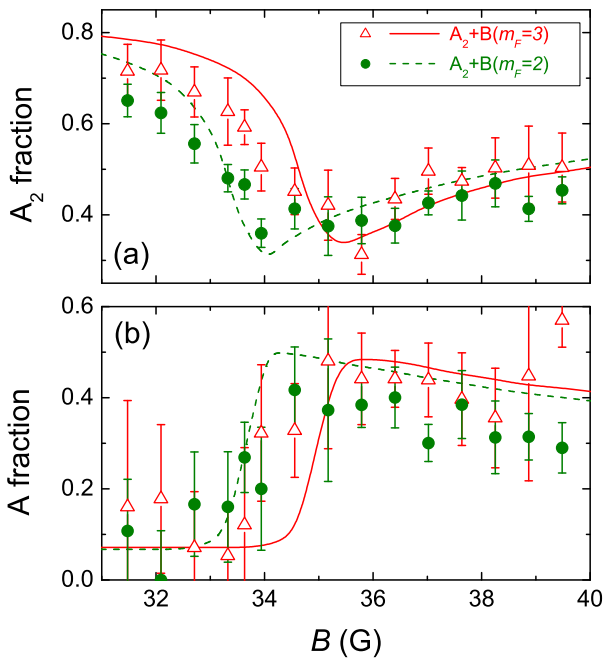}
\caption{(color online) Measurement of the fraction of A$_2$ molecules (a) and A atoms (b), after a fixed storage time of 22\,ms of the B+A$_2$ mixtures at a temperature of 100(10)\,nK. The fractions are defined as the A$_2$ molecule and A atom number normalized to the initial A$_2$ molecule number (about 4000). The data points are averaged over three to five measurement runs and the error bars represent the statistical uncertainty. The lines are obtained from the rate equations using the theoretical results for the total loss rates (a) and exchange rates (b), as shown in Fig.~\ref{lossrates}.}\label{Bscan}
\end{figure}

A ``smoking gun'' for the exchange process is the appearance of atoms in state A. They show up as a reaction product and remain in the trap because of the small energy release. In a second set of experiments, we measure the number of A$_2$ molecules and A atoms in a small magnetic field range around 35\,G after holding the A$_2$+B mixtures for $m_F=2$ and 3 for a fixed storage time. The results are shown in Fig.~\ref{Bscan}(a) and (b), for the fraction of remaining A$_2$ molecules and A atoms, respectively. We indeed observe the onset of the appearance of A atoms around 35\,G, providing clear evidence for the exchange process. Most of lost A$_2$ molecules are accompanied by the appearance of A atoms, indicating that the molecular loss is dominated by the exchange process. The curves in Fig.~\ref{Bscan}(a) and (b) are obtained from the theoretical total loss and exchange rates, respectively. The number of A atoms is consistent with the calculations \cite{atombackground}.

The observation of A atoms implies that also AB molecules are produced. However, we cannot observe trapped AB molecules because of the absence of Feshbach resonances in the present magnetic field region to dissociate the AB molecules \cite{Julienne2008privat}. Furthermore, the AB molecule is expected to very rapidly decay via spin relaxation \cite{Mark2007sou}.

\begin{figure}
\includegraphics[width=8.5cm]{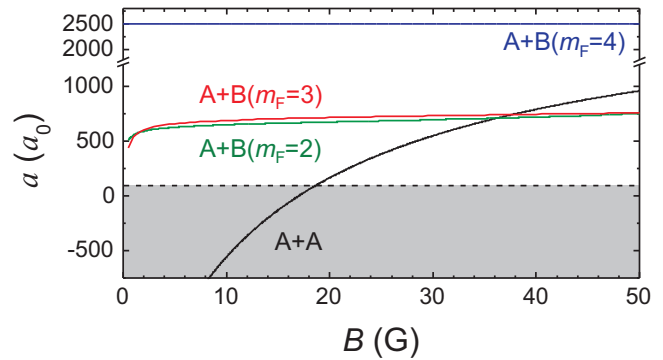}
\caption{(color online) Magnetic field dependence of the two-body scattering lengths for the A+A and the three A+B channels, in units of the Bohr radius $a_0$. The non-shaded region represents the universal regime, which for Cs is realized for $a\gg 100 a_0$ \cite{Kohler2006poc}. Narrow Feshbach resonances resulting from higher partial waves are neglected here.}\label{scatteringlengths}
\end{figure}

A remarkable finding is the dominance of the exchange process over inelastic decay to more deeply bound states. This can be explained by the fact that the exchange process takes place in a regime where the relevant two-body scattering lengths are very large and the dimer A$_2$ and AB are in the quantum halo regime \cite{Braaten2006uif}. The relevant scattering lengths are shown in Fig.~\ref{scatteringlengths}; they are all in the universal regime for $B>20$\,G. Near and above the magnetic fields at which the A$_2$+B and A+AB channels cross, the dominance of the exchange process can be attributed to the large wavefunction overlap between these two channels resulting from the large and similar extent of the A$_2$ and AB states. That is in contrast to the small overlap between the A$_2$+B channel to more deeply bound channels because of the small extent of the deeply bound states. In our calculations we were able to verify that, when the channel crossing occurs at small scattering lengths, all processes have the same importance and the exchange process is not favored.

Our theoretical analysis also indicates that for large scattering lengths the exchange process occurs predominantly when all three atoms are within distances comparable to the scattering lengths and, therefore, none of the atoms ``see'' the short-range details of the interatomic interactions. In contrast, atom-molecule collisions that populate more deeply bound states require all three atoms to approach to short distances. Therefore, our theoretical findings suggest that the dominance of the exchange process is a consequence of the universal regime of halo dimers. In fact, the nearly equal experimental rates for the $m_F=2$ and $m_F=3$ cases are consistent with our expectation that, near the channel crossing, the main collisional behavior depends only on the scattering lengths. Therefore we also expect atom exchange to be the dominant process in heteronuclear systems when both the heteronuclear and one of the homonuclear scattering lengths are large and positive. The possible high efficiency of atom exchange for halo molecules was already pointed out long time ago \cite{Stwalley1978lrm}.

Finally we turn to our observations for the $m_F=4$ case (see Fig.~\ref{lossrates}). Here we observe no resonance, but instead very small loss rates for $B>20$\,G. We find that this is well reproduced by the model calculations. Although the A$_{2}$+B and A+AB channels do not cross on the present range of magnetic fields, the fact that both scattering lengths are large does affect the collisional behavior. In our numerical calculations we find a strong coupling between between the A$_{2}$+B and A+AB channels at interatomic separations comparable to the A+A scattering length. Such a coupling manifests itself in the appearance of a repulsive barrier in both channels, leading to the observed suppression of A$_{2}$+B collisions \cite{DIncao2009}.

To summarize, we have observed an exchange process in an optically trapped ultracold sample of atoms and Feshbach molecules. In a mixture of A$_2$ molecules and B atoms, where A and B are two hyperfine sublevels of Cs, we observe the appearance of free atoms in state A once the exchange process is magnetically tuned to be exoergic. This lead to a resonant enhancement near threshold. In contrast to relaxation to more deeply bound molecular states, the exchange process does not lead to trap loss. The magnetic field dependence of the measured total inelastic collision and exchange rates are in excellent agreement with model calculations. The high efficiency of the exchange process is explained by the halo character of both the initial and final molecular states.

Our observation represents an elementary example for the possibility to control a reaction that involves ultracold molecules. Since field-dependent resonance and threshold phenomena are ubiquitous in molecular gases, we expect that similar possibilities will arise in many situations and offer intriguing possibilities in the developing field of ultracold molecular quantum gases and controlled ultracold chemistry.

We thank P. S. Julienne for helpful discussions and providing us the scattering length calculations. We acknowledge support by the Austrian Science Fund (FWF) within SFB 15 (project part 16). S.~K.\ was supported within the Marie Curie Intra-European Program of the European Commission. F.~F.\ was supported within the Lise Meitner program of the FWF. J.~P.~D.\ and B.~D.~E.\ acknowledge support from U.~S.\ NSF.


\begin{thebibliography}{25}
\expandafter\ifx\csname natexlab\endcsname\relax\def\natexlab#1{#1}\fi
\expandafter\ifx\csname bibnamefont\endcsname\relax
  \def\bibnamefont#1{#1}\fi
\expandafter\ifx\csname bibfnamefont\endcsname\relax
  \def\bibfnamefont#1{#1}\fi
\expandafter\ifx\csname citenamefont\endcsname\relax
  \def\citenamefont#1{#1}\fi
\expandafter\ifx\csname url\endcsname\relax
  \def\url#1{\texttt{#1}}\fi
\expandafter\ifx\csname urlprefix\endcsname\relax\def\urlprefix{URL }\fi
\providecommand{\bibinfo}[2]{#2}
\providecommand{\eprint}[2][]{\url{#2}}

 \bibitem[{\citenamefont{Krems et~al.}(2009)\citenamefont{Krems, Friedrich,
  and Stwalley}}]{Krems2009cmt}
\bibinfo{author}{\bibfnamefont{R.~V.}~\bibnamefont{Krems}},
  \bibinfo{author}{\bibfnamefont{B.}~\bibnamefont{Friedrich}}, \bibnamefont{and}
  \bibinfo{author}{\bibfnamefont{W.~C.}~\bibnamefont{Stwalley}}, \bibnamefont{eds.}
  \emph{\bibinfo{title}{Cold Molecules: Theory, Experiment, Applications}}
  (\bibinfo{publisher}{Taylor \& Francis}, \bibinfo{year}{2009}).

  \bibitem[{\citenamefont{Danzl et~al.}(2008)\citenamefont{Danzl, Haller,
  Gustavsson, Mark, Hart, Bouloufa, Dulieu, Ritsch, and
  N\"{a}gerl}}]{Danzl2008qgo}
\bibinfo{author}{\bibfnamefont{J.~G.} \bibnamefont{Danzl}} {\it et al.},
  \bibinfo{journal}{Science} \textbf{\bibinfo{volume}{321}},
  \bibinfo{pages}{1062} (\bibinfo{year}{2008});
  \bibinfo{author}{\bibfnamefont{J.}~\bibnamefont{Deiglmayr}} {\it et al.},
  \bibinfo{journal}{Phys.\ Rev.\ Lett.} \textbf{\bibinfo{volume}{101}},
  \bibinfo{pages}{133004} (\bibinfo{year}{2008});
\bibinfo{author}{\bibfnamefont{F.}~\bibnamefont{Lang}} {\it et al.},
  {\it ibid.} \textbf{\bibinfo{volume}{101}},
  \bibinfo{pages}{133005} (\bibinfo{year}{2008});
\bibinfo{author}{\bibfnamefont{K.-K.} \bibnamefont{Ni}} {\it et al.},
  \bibinfo{journal}{Science} \textbf{\bibinfo{volume}{322}},
  \bibinfo{pages}{231} (\bibinfo{year}{2008}); \bibinfo{author}{\bibfnamefont{J.~G.} \bibnamefont{Danzl}} {\it et al.}, \bibinfo{note}{arXiv:0909.4700}.



\bibitem[{\citenamefont{Carr}(2009)}]{Carr2009cau}
\bibinfo{author}{\bibfnamefont{L.~D.} \bibnamefont{Carr}},
\bibinfo{author}{\bibfnamefont{D.} \bibnamefont{DeMille}},
\bibinfo{author}{\bibfnamefont{R.~V.} \bibnamefont{Krems}}, \bibnamefont{and}
\bibinfo{author}{\bibfnamefont{J.} \bibnamefont{Ye}},
\bibinfo{journal}{New.\ J.\ Phys.} \textbf{\bibinfo{volume}{11}},
\bibinfo{pages}{055049} (\bibinfo{year}{2009}).

\bibitem[{\citenamefont{Krems}(2008)}]{Krems2008ccc}
\bibinfo{author}{\bibfnamefont{R.~V.} \bibnamefont{Krems}},
\bibinfo{journal}{Phys.\ Chem.\ Chem.\ Phys.} \textbf{\bibinfo{volume}{10}},
\bibinfo{pages}{4079} (\bibinfo{year}{2008}).

\bibitem[{\citenamefont{Strecker et~al.}(2003)\citenamefont{Strecker,
 Partridge, and Hulet}}]{Strecker2003coa}
\bibinfo{author}{\bibfnamefont{K.~E.} \bibnamefont{Strecker}},
  \bibinfo{author}{\bibfnamefont{G.~B.} \bibnamefont{Partridge}},
  \bibnamefont{and} \bibinfo{author}{\bibfnamefont{R.~G.} \bibnamefont{Hulet}},
  \bibinfo{journal}{Phys.\ Rev.\ Lett.} \textbf{\bibinfo{volume}{91}},
  \bibinfo{eid}{080406} (\bibinfo{year}{2003});
%\bibitem[{\citenamefont{Cubizolles et~al.}(2003)\citenamefont{Cubizolles,
%Bourdel, Kokkelmans, Shlyapnikov, and Salomon}}]{Cubizolles2003pol}
\bibinfo{author}{\bibfnamefont{J.}~\bibnamefont{Cubizolles}},  {\it et al.},
{\it ibid.} \textbf{\bibinfo{volume}{91}},
\bibinfo{pages}{240401} (\bibinfo{year}{2003});
%\bibitem[{\citenamefont{Jochim et~al.}(2003)\citenamefont{Jochim, Bartenstein, Altmeyer, % Hendl, Chin, {Hecker Denschlag}, and Grimm}}]{Jochim2003pgo}
\bibinfo{author}{\bibfnamefont{S.}~\bibnamefont{Jochim}},  {\it et al.},
{\it ibid.} \textbf{\bibinfo{volume}{91}},
\bibinfo{pages}{240402} (\bibinfo{year}{2003}).

\bibitem[{\citenamefont{Knoop et~al.}(2009)\citenamefont{Knoop, Ferlaino, Mark,
  Berninger, Sch\"{o}bel, N\"{a}gerl, and Grimm}}]{Knoop2009ooa}
\bibinfo{author}{\bibfnamefont{S.}~\bibnamefont{Knoop}} {\it et al.},
  \bibinfo{journal}{Nature Phys.} \textbf{\bibinfo{volume}{5}},
  \bibinfo{pages}{227} (\bibinfo{year}{2009}).

\bibitem[{\citenamefont{Chin et~al.}(2005)\citenamefont{C. Chin and T. Kraemer and M. Mark and J.
Herbig and P. Waldburger and H.-C. N\"agerl and R. Grimm}}]{Chin2005oof}
\bibinfo{author}{\bibfnamefont{C.} \bibnamefont{Chin}} {\it et al.},
  \bibinfo{journal}{Phys.\ Rev.\ Lett.} \textbf{\bibinfo{volume}{94}},
  \bibinfo{eid}{123201} (\bibinfo{year}{2005});
%\bibitem[{\citenamefont{Ferlaino et~al.}(2009)\citenamefont{}}]{Ferlaino2009cpo}
\bibinfo{author}{\bibfnamefont{F.}~\bibnamefont{Ferlaino}} {\it et al.},
  \bibinfo{journal}{Laser Phys., in press},  \bibinfo{note}{arXiv:0904.0935}.

\bibitem[{\citenamefont{Mukaiyama et~al.}(2004)\citenamefont{Mukaiyama,
Abo-Shaeer, Xu, Chin, and Ketterle}}]{Mukaiyama2004dad}
\bibinfo{author}{\bibfnamefont{T.}~\bibnamefont{Mukaiyama}} {\it et al.},
\bibinfo{journal}{Phys.\ Rev.\ Lett.} \textbf{\bibinfo{volume}{92}},
\bibinfo{pages}{180402} (\bibinfo{year}{2004});
%\bibitem[{\citenamefont{Staanum et~al.}(2006)\citenamefont{Staanum, Kraft,
%Lange, Wester, and Weidem{\"u}ller}}]{Staanum2006eio}
\bibinfo{author}{\bibfnamefont{P.}~\bibnamefont{Staanum}} {\it et al.},
{\it ibid.} \textbf{\bibinfo{volume}{96}},
\bibinfo{eid}{023201} (\bibinfo{year}{2006});
%\bibitem[{\citenamefont{Zahzam et~al.}(2006)\citenamefont{Zahzam, Vogt,
%  Mudrich, Comparat, and Pillet}}]{Zahzam2006amc}
\bibinfo{author}{\bibfnamefont{N.}~\bibnamefont{Zahzam}} {\it et al.},
{\it ibid.} \textbf{\bibinfo{volume}{96}},
\bibinfo{eid}{023202} (\bibinfo{year}{2006});
 \bibinfo{author}{\bibfnamefont{N.}~\bibnamefont{Syassen}} {\it et al.},
\bibinfo{journal}{Phys.\ Rev.\ A} \textbf{\bibinfo{volume}{74}},
\bibinfo{eid}{062706} (\bibinfo{year}{2006});
%\bibitem[{\citenamefont{Zirbel et~al.}(2008)\citenamefont{Zirbel, Ni,
%  Ospelkaus, D'Incao, Wieman, Ye, and Jin}}]{Zirbel2008cso}
\bibinfo{author}{\bibfnamefont{J.~J.} \bibnamefont{Zirbel}} {\it et al.},
\bibinfo{journal}{Phys.\ Rev.\ Lett.} \textbf{\bibinfo{volume}{100}},
  \bibinfo{pages}{143201} (\bibinfo{year}{2008});
%\bibitem[{\citenamefont{Hudson et~al.}(2008)\citenamefont{Hudson, Gilfoy,
%  Kotochigova, Sage, and DeMille}}]{Hudson2008ico}
\bibinfo{author}{\bibfnamefont{E.~R.} \bibnamefont{Hudson}} {\it et al.},
  {\it ibid.} \textbf{\bibinfo{volume}{100}},
  \bibinfo{pages}{203201} (\bibinfo{year}{2008}).

\bibitem[{\citenamefont{Suno et~al.}(2002)\citenamefont{Suno, Esry, Greene, and
 Burke}}]{Suno2002tbr}
\bibinfo{author}{\bibfnamefont{H.}~\bibnamefont{Suno}},
\bibinfo{author}{\bibfnamefont{B.~D.} \bibnamefont{Esry}},
\bibinfo{author}{\bibfnamefont{C.~H.} \bibnamefont{Greene}},
\bibnamefont{and} \bibinfo{author}{\bibfnamefont{J.~P.} \bibnamefont{Burke}},
\bibinfo{journal}{Phys.\ Rev.\ A} \textbf{\bibinfo{volume}{65}},
\bibinfo{eid}{042725} (\bibinfo{year}{2002}).

\bibitem[{\citenamefont{K\"ohler et~al.}(2006)\citenamefont{K\"ohler, G\'oral,
  and Julienne}}]{Kohler2006poc}
\bibinfo{author}{\bibfnamefont{T.}~\bibnamefont{K\"ohler}},
  \bibinfo{author}{\bibfnamefont{K.}~\bibnamefont{G\'oral}}, \bibnamefont{and}
  \bibinfo{author}{\bibfnamefont{P.~S.} \bibnamefont{Julienne}},
  \bibinfo{journal}{Rev.\ Mod.\ Phys.} \textbf{\bibinfo{volume}{78}},
  \bibinfo{eid}{1311} (\bibinfo{year}{2006}).

 \bibitem[{\citenamefont{Ferlaino et~al.}(2009)\citenamefont{Ferlaino, Knoop,
  and Grimm}}]{Ferlaino2009ufm}
\bibinfo{author}{\bibfnamefont{F.}~\bibnamefont{Ferlaino}},
  \bibinfo{author}{\bibfnamefont{S.}~\bibnamefont{Knoop}}, \bibnamefont{and}
  \bibinfo{author}{\bibfnamefont{R.}~\bibnamefont{Grimm}}, \bibnamefont{in Ref.\,\cite{Krems2009cmt}}.

\bibitem[{\citenamefont{Braaten et~al.}(2006)\citenamefont{Braaten, Hammer}}]{Braaten2006uif}
\bibinfo{author}{\bibfnamefont{E.}~\bibnamefont{Braaten}}
\bibnamefont{and} \bibinfo{author}{\bibfnamefont{H.~W.} \bibnamefont{Hammer}},
\bibinfo{journal}{Phys.\ Rep.} \textbf{\bibinfo{volume}{428}},
\bibinfo{eid}{259} (\bibinfo{year}{2006}).

\bibitem[{\citenamefont{Chin et~al.}(2004)\citenamefont{Chin, Vuleti\'c, Kerman, Chu, Tiesinga, Leo, and Williams}}]{Chin2004pfs}
\bibinfo{author}{\bibfnamefont{C.}~\bibnamefont{Chin}} {\it et al.},
  \bibinfo{journal}{Phys.\ Rev.\ A} \textbf{\bibinfo{volume}{70}},
  \bibinfo{pages}{032701} (\bibinfo{year}{2004}).

\bibitem[{\citenamefont{Mark et~al.}(2007)\citenamefont{Mark, Ferlaino, Knoop, Danzl, Kraemer, Chin, N\"{a}gerl, and Grimm}}]{Mark2007sou}
\bibinfo{author}{\bibfnamefont{M.}~\bibnamefont{Mark}} {\it et al.},
  \bibinfo{journal}{Phys.\ Rev.\ A} \textbf{\bibinfo{volume}{76}},
  \bibinfo{pages}{042514} (\bibinfo{year}{2007}).

\bibitem[{\citenamefont{Julienne and Tiesinga}(2008)}]{Julienne2008privat}
\bibinfo{author}{\bibfnamefont{P.~S.} \bibnamefont{Julienne}} \bibnamefont{and}
\bibinfo{author}{\bibfnamefont{E.}~\bibnamefont{Tiesinga}},
\bibinfo{note}{privat communication}.

\bibitem[{\citenamefont{Gao}(2004)}]{Gao2004bea}
\bibinfo{author}{\bibfnamefont{B.}~\bibnamefont{Gao}}, \bibinfo{journal}{J.\
Phys.\ B: At.\ Mol.\ Opt.\ Phys.} \textbf{\bibinfo{volume}{37}},
\bibinfo{pages}{4273} (\bibinfo{year}{2004}).

\bibitem[{\citenamefont{Ferlaino et~al.}(2008)\citenamefont{Ferlaino, Knoop,
Mark, Berninger, Sch\"{o}bel, N\"{a}gerl, and Grimm}}]{Ferlaino2008cbt}
\bibinfo{author}{\bibfnamefont{F.}~\bibnamefont{Ferlaino}} {\it et al.},
\bibinfo{journal}{Phys.\ Rev.\ Lett.} \textbf{\bibinfo{volume}{101}},
\bibinfo{pages}{023201} (\bibinfo{year}{2008}).

\bibitem[{\citenamefont{atombackground}(000)}]{atombackground}
\bibinfo{note}{We attribute the small amount of A atoms present at lower magnetic field values to the non-perfect MW transfer. However, because the MW transfer is done at a fixed magnetic field value, the number of non-transferred A atoms is independent of the magnetic field at which the measurement is performed. Therefore we expect a constant background, which is added to the theoretical results to obtain the curves of Fig.~\ref{Bscan}(b).}

\bibitem[{\citenamefont{Stwalley}(1978)}]{Stwalley1978lrm}
\bibinfo{author}{\bibfnamefont{W.~C.} \bibnamefont{Stwalley}},
\bibinfo{journal}{Contemp.\ Phys.} \textbf{\bibinfo{volume}{19}},
\bibinfo{pages}{65} (\bibinfo{year}{1978});
%\bibitem[{\citenamefont{Stwalley}(2004)}]{Stwalley2004car}
\bibinfo{author}{\bibfnamefont{W.~C.} \bibnamefont{Stwalley}},
\bibinfo{journal}{Can.\ J.\ Chem.} \textbf{\bibinfo{volume}{82}},
\bibinfo{pages}{709} (\bibinfo{year}{2004}).

\bibitem[{\citenamefont{D'Incao and Esry}(2009)}]{DIncao2009}
\bibinfo{author}{\bibfnamefont{J.~P.} \bibnamefont{D'Incao}} \bibnamefont{and}
\bibinfo{author}{\bibfnamefont{B.~D.} \bibnamefont{Esry}}, \bibinfo{note}{in
preparation}.

\end{thebibliography}
\end{document}